\documentclass[prb,showpacs,twocolumn,amsmath]{revtex4-1}

\usepackage{ulem}
\usepackage{color}
\usepackage[dvipdfm]{graphicx}
\usepackage{graphicx}
\usepackage{dcolumn}
\usepackage{bm}
\usepackage{times}

\begin{document}

\preprint{penetration depth}

\title{Effect of magnetic criticality and Fermi-surface topology on the magnetic penetration depth}%

\author{Takuya Nomoto}%
\email{nomoto.takuya@scphys.kyoto-u.ac.jp}
\author{Hiroaki Ikeda}%
\affiliation{Department of Physics, Kyoto University, Kyoto, 606-8502, Japan}%

\date{\today}%

\begin{abstract}
We investigate the effect of anti-ferromagnetic (AF) quantum criticality on the magnetic penetration depth $\lambda(T)$ in line-nodal superconductors, including the cuprates, the iron pnictides, and the heavy-fermion superconductors. The critical magnetic fluctuation renormalizes the current vertex and drastically enhances zero-temperature penetration depth $\lambda(0)$, which is more remarkable in the iron-pnictide case due to the Fermi-surface topology. Additional temperature ($T$) dependence of the current renormalization makes  the expected $T$-linear behavior at low temperatures approaching to $T^{1.5}$ asymptotically. These anomalous behaviors are well consistent with experimental observations. We stress that $\lambda(T)$ is a good probe to detect the AF quantum critical point in the superconducting state. 
\end{abstract}


\maketitle

The discovery of high-$T_c$ superconductivity in iron-pnictides has demonstrated possible existence of high-$T_c$ materials other than the cuprates, and continues to promote currently intensive research studies. It is thought that high transition temperatures in these materials cannot be explained by the conventional phonon-mediated mechanism. The underlying pairing mechanism is one of most interesting and highly debated subjects in modern condensed-matter physics. A plausible scenario of unconventional pairing mechanism is magnetic-fluctuations mediated superconductivity. Indeed these materials share similar phase diagrams; unconventional superconductivity appears in a close proximity to the anti-ferromagnetic (AF) phase boundary. However, the details are dependent on each material, and then it is still not so clear whether the pairing mechanism in these materials can be understood on the same footing or not. 

A systematic study in BaFe$_2$(As$_{1-x}$P$_x$)$_2$ system and its comparison with the other systems give us a good opportunity to elucidate a relation between AF fluctuations and  high transition temperatures~\cite{rf:Jiang2009, rf:Nakai2010, rf:Kasahara2010, rf:Shishido2010, rf:Hashimoto2012}. In the normal state, the NMR relaxation rate $1/T_1T$ increases on cooling~\cite{rf:Nakai2010}, the electric resistivity shows $T$-linear behavior~\cite{rf:Kasahara2010} and the de Haas-van Alphen (dHvA) measurement revealed the enhanced effective mass toward the AF phase boundary~\cite{rf:Shishido2010}. These all observations imply that the AF quantum critical point (AF QCP) can be located at around the optimal doping $x=0.3$ with the highest $T_c$. Recently, in the superconducting state, a sharp peak of the zero-temperature penetration depth $\lambda(0)$ has been observed at the critical doping, indicative of the AF QCP beneath the superconducting dome~\cite{rf:Hashimoto2012}. From the conventional formula $\lambda^2=m^*c^2/(4\pi e^2 n)$ [\onlinecite{rf:Prozorov2006}], this means strong enhancement of the effective mass $m^*$ with finite carrier density $n$ [\onlinecite{rf:Varma1986}]. Such a trend of the increase toward AF phase boundary has been also reported in a heavy fermion compound CeCoIn$_5$ [\onlinecite{rf:Howald2013}]. Moreover, concerning $T$ dependence, remarkable deviation from the expected $T$-linear behavior, rather $T^{1.5}$ dependence has been reported in these line-nodal superconductors~\cite{rf:Chia2003, rf:Ozcan2003, rf:Hashimoto2013}, as well as organic compounds $\kappa-$(BEDT-TTF)$_2X$ ($X$=Cu[N(CN)$_2$]Br and Cu(NCS)$_2$) [\onlinecite{rf:Carrington1999}]. The London penetration depth $\lambda(T)$ is sensitive to quasi-particle low-energy excitations in the superconducting state. Generally, its $T$ dependence at low temperatures shows the exponential decay in fully-gapped superconductors, and $T$-linear dependence in line-nodal $d$-wave superconductors~\cite{rf:Prozorov2006}. Experimentally, however, it is difficult to find $T$-linear dependence at low $T$ limit due to small amount of impurities and/or lattice imperfection. Empirically, such deviation from $T$-linear can be well described by interpolation formula $T^2/(T+T^*)$ with $T^*$, where $T^*$ is $T_{\rm imp}^* \approx 0.83 \sqrt{\Gamma \Delta_0}$ with an impurity scattering rate $\Gamma$ and a maximum gap $\Delta_0$ for the impurity scattering origin~\cite{rf:Hirschfeld1993}, or $T_{\rm loc}^* \approx \Delta_0 \xi(0) / \lambda(0)$ with a coherent length $\xi(0)$ for the non-locality effect~\cite{rf:Kosztin1997}. However, the observed $T^{1.5}$ behavior means too much large $T^*$ [\onlinecite{rf:Carrington1999, rf:Hashimoto2013}]. Instead, a possible scenario independent of $T^*$, the effect of additional momentum dependence of mass renormalization below $T_c$, has been discussed~\cite{rf:Hashimoto2013}.

In this letter, we investigate the effect of AF quantum criticality on $\lambda(T)$ based on the standard Fermi liquid formula, which was developed in Ref. [\onlinecite{rf:Leggett1965,rf:Larkin1964,rf:Jujo2000,rf:Jujo2002}]. The central physical quantities are the mass renormalization $m/m^*$ and the current vertex ${\bm j}^*({\bm k})$. 
The latter
includes the Fermi-liquid type backflow effect, corresponding to $1+F^s_1/3$ in the isotropic Fermi-liquid system, which cancels out $m/m^*$ in the Galilean invariant system due to the Ward identity~\cite{rf:Leggett1965,rf:Larkin1964,rf:Jujo2000}. We here consider a two-band model corresponding to the iron pnictides, and a single band model in cuprates to mimic heavy fermion superconductors. In these lattice systems, $m/m^*$ is not necessarily related to the Fermi liquid corrections, and then can be strongly renormalized as approaching to AF-QCP. Moreover, since the current vertex is also reduced, $\lambda^2(0)$ can be more enhanced than $m^*/m$. Interestingly, we find that the enhancement of $\lambda(0)$ is more remarkable in the iron-pnictide case due to the Fermi-surface topology, which can explain dramatically sharp peak observed only in the iron-pnictides. In addition, concerning $T$ dependence of $\lambda(T)$ in line-nodal superconductors, we find that the current vertex correction provides additional $T$ dependence in $\lambda(T)$ near the AF QCP, which approaches to $T^{1.5}$ asymptotically rather than the expected $T$-linear dependence. 
We here show generic features of $\lambda(T)$ expected in line-nodal superconductors located just close to AF-QCP beneath the superconducting dome.

{\it Formalism ---} First of all, we start with a brief summary of a theoretical approach of $\lambda(T)$ based on the Fermi liquid theory in superconducting states, following in Ref.[\onlinecite{rf:Jujo2000}].
For $x$ direction,
\begin{align}
\frac{1}{\lambda_{xx}^2(T)} \propto \int_{FS}\frac{dS_{\bm k}}{(2\pi)^2 |{\bm v}^*({\bm k})|}v_x^*(\bm k)(1-Y({\bm k};T)) \bar{v}^*_x({\bm k};T), \label{lambda}
\end{align}
where the integral is performed on the Fermi surface, and $Y({\bm k};T)$ is the so-called Yosida function, which decreases smoothly from $Y({\bm k};T_c)=1$ and vanishes at $T=0$, and then dominates $T$ dependence of $\lambda(T)$ at low $T$ limit, which gives rise to $T$-linear behavior in line-nodal superconductors. ${\bm v}^*({\bm k})$ is a renormalized quasi-particle velocity, ${\bm v}^*({\bm k}) =z_{\bm k} \frac{\partial}{\partial{\bm k}}\left(\varepsilon_{\bm k}+{\rm Re}\Sigma^R({\bm k},0)\right)$
with a band dispersion $\varepsilon_{\bm k}$, a self-energy shift ${\rm Re}\Sigma^R({\bm k},0)$, and a mass renormalization factor,
\begin{align}
z_{\bm k}=\left( 1-\frac{\partial {\rm Re}\Sigma^R({\bm k},\omega)}
{\partial\omega}\biggr|_{\omega=0} \right)^{-1}.
\end{align}
The renormalized current vertex, $\bar{\bm v}^*({\bm k};T)$, is defined by
\begin{align}
\bar{\bm v}^*({\bm k};T) ={\bm j}^*({\bm k})-\int_{FS} \frac{dS_{\bm k'}}{(2\pi)^2|{\bm v}^*({\bm k}')|} f_{{\bm k},{\bm k}'}Y({\bm k}';T)  \bar{\bm v}^*({\bm k'};T), \label{currentT}
\end{align}
with the quasi-particle current density,
\begin{align}
{\bm j}^*({\bm k}) ={\bm v}^*({\bm k})+\sum_{{\bm k}'}f_{{\bm k},{\bm k}'}\left(-\frac{\partial f(\varepsilon_{\bm k'})}{\partial \varepsilon_{\bm k'}}\right){\bm v}^*({\bm k'}), \label{current}
\end{align}
where $f(\varepsilon)$ is a Fermi distribution function, and $f_{{\bm k},{\bm k}'}$ an effective interaction between quasi-particles. 
Note that $\bar{\bm v}^*({\bm k};T)$ includes a repetition of $Y({\bm k};T)$, which can produce additional $T$ dependence in nodal superconductors.
Following these formula, zero-temperature penetration depth $\lambda(0)$ is simply given by 
\begin{align}
\frac{1}{\lambda_{xx}^2(0)} \propto \int_{FS}\frac{dS_{\bm k}}{(2\pi)^2 |{\bm v}^*({\bm k})|}v_x^*(\bm k) j_x^*({\bm k}). \label{lambda0}
\end{align}
Here $j_x^*({\bm k})$ includes both the mass renormalization $m/m^*$ and the current vertex due to the backflow effect~\cite{rf:note01}. In the Galilean invariant isotropic systems, the current density is reduced to the unrenormalized velocity ${\bm v}({\bm k})={\bm v}^*({\bm k})/z_k$ due to the above-mentioned Ward identity, which is identical to the $unrenormalized$ penetration depth $\sim n/m$. In generic lattice system, such compensation does not work. If neglecting the backflow effect as the zeroth-order approximation, i.e., including only the effect of mass renormalization, then the phenomenological $renormalized$ value $\sim n/m^*$ is obtained, which is below referred to as ``without the current vertex corrections (w/o CVC)''. 

In this letter, to investigate the effect of AF critical fluctuations, we consider the quasi-particle interactions $f_{\bm{k,k'}}=z_{\bm k}\Gamma_{\bm{k,k'}}(\omega=0)z_{\bm k'}$ with $\Gamma_{\bm{k,k'}}(\omega)=\alpha U^2\chi(\bm{k-k'},\omega)$ [\onlinecite{rf:note02}]. In the normal state, the strong AF spin fluctuations $\chi(\bm{q},\omega)$ is defined as follows, based on the self-consistent renormalization theory~\cite{rf:Moriya2000}. With the AF wave vector, ${\bm Q}=(\pi,\pi)$,
\begin{align}
\chi({\bm Q}+{\bm q},\omega) = \frac{\chi({\bm Q}+{\bm q})}{1-i\omega/\Gamma_{{\bm Q}+{\bm q}}},
\end{align}
where $\chi^{-1}({\bm Q}+{\bm q})=\chi^{-1}({\bm Q})+Aq^2=\eta+Aq^2$ and $\Gamma_{{\bm Q}+{\bm q}}=\Gamma(\kappa^2+q^2 )$. $A$ and $\Gamma$ are material dependent parameters, related to characteristic temperatures $T_A$ and $T_0$ via $T_A = Aq_B^2/2$ and $T_0 = \Gamma q_B^2/2\pi$ with $q_B$, a cut off wave vector. $T_A$ and $T_0$ represent the extent of the AF spin fluctuations in momentum space and energy space, respectively~\cite{rf:note03}.
Since $A\chi({\bm Q})$ is a square of magnetic correlation length $\xi^2(T)$, $\eta$ represents a distance to the AF QCP at $\eta=0$.
The preceding studies show that the contribution at one-loop level can well describe non-Fermi liquid behavior near the AF QCP.
Generally, in the superconducting state, the low-energy excitations of the spin fluctuations should be drastically suppressed due to the superconducting gap formation. 
However, near the AF QCP in the superconducting state, the low-energy excitations should increase again because the AF QCP is a point at which a soft spin-excitation mode touch zero energy at ${\bm Q}$ [\onlinecite{rf:Takimoto2002}].
In the present calculations, the key parameters are the effective interaction $f_{{\bm k},{\bm k}'}$ and the mass renormalization factor $z_k$. Although these might have strong $T$ dependence for $T<T_c$, only weak $T$ dependence has been observed at least within the fluctuation-exchange approximations (FLEX)~\cite{rf:Jujo2002}. Thus we here do not consider $T$ dependence of these key parameters for simplicity.
This assumption allows us to evaluate the penetration depth with the same effective interaction as that in the normal state. It has an advantage of the use of $T_A$ and $T_0$ obtained by experimental data. The self-energy is evaluated at one-loop level, $\Sigma(\bm{k},i\omega_n)=\sum_{\bm{k},n'}\Gamma_{\bm{k,k'}}(i\omega_n-i\omega_n')G(\bm{k'},i\omega_n')$ with Fermion Matsubara frequencies $\omega_n=(2n+1)\pi T$. The mass renormalization factor of Eq.(2) is approximated as
\begin{align}
z_{\bm k}\simeq \left( 1-\frac{{\rm Im}\Sigma({\bm k},i\pi T)}{i\pi T} \right)^{-1},
\end{align}
at $T=T_c$.
Hereafter, let us discuss the obtained results~\cite{rf:Levchenko2012}.

\begin{figure}[h]
\centering
\includegraphics[width=60mm]{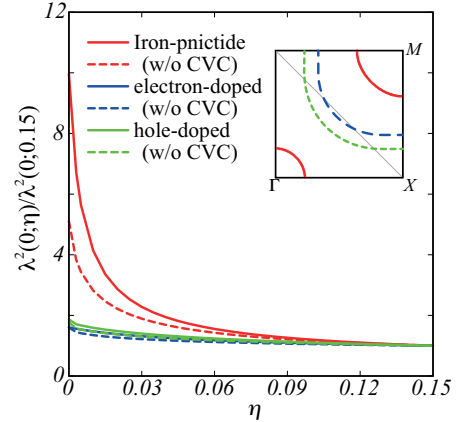}
\caption{(Color online) Zero-temperature penetration depth $\lambda^2(0)$ as a function of $\eta$ in a unit of $t$, which denotes a distance to the AF QCP. In the iron-pnictide case, a significant enhancement is observed both with and without the current vertex corrections. In the inset, red solid, blue dashed, green dotted lines, respectively, correspond to the Fermi surfaces in the iron-pnictide case, and the electron- and the hole-doped cuprate cases.}
\end{figure}
{\it Zero-temperature penetration depth ---}
In Fig.1, we first demonstrate $\eta$ dependence of $\lambda^2(0)$ for three cases; the iron-pnictide case, the electron- and the hole-doped cuprate cases~\cite{rf:note04}. The inset is the corresponding Fermi surfaces. We find a dramatic enhancement of $\lambda^2(0)$ in the iron-pnictide case, as compared with the cuprate cases. 
\begin{figure}[h]
\centering
\includegraphics[width=80mm]{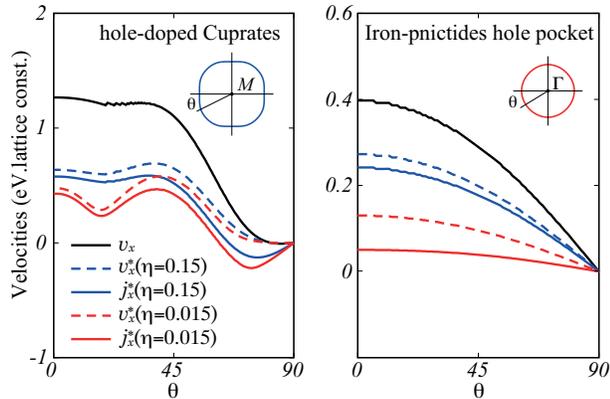}
\caption{(Color online) Angle ($\theta$) dependence of the $x$ component of the Fermi velocity $v_x({\bm k})$, the renormalized velocity $v_x^*({\bm k})$, and the renormalized current $j_x^*({\bm k})$ in the hole-doped cuprate case (Left), and in the hole pocket around $\Gamma$ point of the iron-pnictide case (Right). The minima observed in $j_x^*(\bm{k})$ of the left figure corresponds to the hot spots.}
\end{figure}
Even without the current vertex corrections (w/o CVC in Fig.1), similar enhancement is obtained, although the magnitude is clearly suppressed. This result is well consistent with experimental observations; a trend of enhancement toward the AF phase boundary in the cuprates and several heavy-fermion compounds, and more remarkable peak structure observed in BaFe$_2$(As$_{1-x}$P$_x$)$_2$. It indicates that the contribution of the critical fluctuation on $\lambda^2(0)$ is significantly important in the iron pnictides. Magnitude of the contribution is material-dependent, related to a balance between the extent of spin fluctuations in momentum space $\sqrt{\eta/A}$ and the Fermi-surface topology, especially, the size of the Fermi wave vector $k_F$. This is shown more clearly in the angle-resolved $v^*_x({\bm k})$ and $j^*_x({\bm k})$, as illustrated in Fig.2. In the cuprate case, these have strong angle dependence, and strong suppression with the extent of $\sqrt{\eta/A}$ at around $20$ and $70$ degrees. These are the so-called hot spots \cite{rf:Pines1996,rf:Kontani2008}, which are intersections between the Fermi surface and AF Brillouin zone. In this regard, a flat angle dependence in the iron pnictides is indicative that the whole Fermi surface is just like hot spots. Indeed in this case $\sqrt{\eta/A}$ is large enough to cover the Fermi surface. Furthermore, we find that $j^*_x$ is more suppressed than $v^*_x$. This sizable CVC implies the importance of the backflow effect on $\lambda(0)$. In the vicinity of AF QCP, the renormalized current ${\bm j}^*({\bm k})$ is approximated by ${\bm j}^*({\bm k})\approx{\bm v}^*({\bm k})+c {\bm v}^*({\bm k}+{\bm Q})$ with a positive constant $c$. With the opposite sign of ${\bm v}^*({\bm k})$ and ${\bm v}^*({\bm k}+{\bm Q})$, ${\bm j}^*({\bm k})$ and also $\lambda(0)^{-2}$ are always suppressed near the AF QCP. 

\begin{figure}[h]
\centering
\includegraphics[width=60mm]{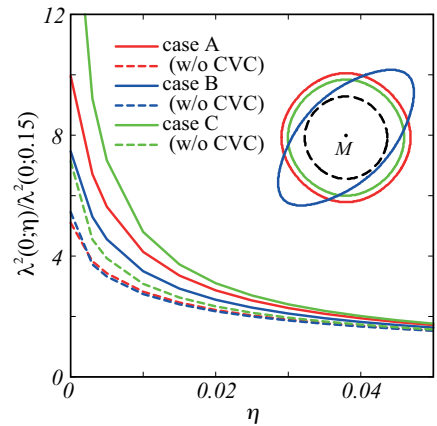}
\caption{(Color online) Effects of the Fermi-surface topology on $\lambda(0)$ in the iron-pnictide case. The corresponding electron Fermi sheets are illustrated in the inset. Case A is identical to the iron-pnictide case in Fig.1. We set $(t'',V,\mu)=(2.0t,4.6t,-1.1t)$ in case B, and $(0,4.8t,-1.3t)$ in case C. In the anisotropic Fermi surface (case B), the contribution of the CVC, the difference between solid and dashed lines, is much suppressed.}
\end{figure}

Here, to clarify the role of the Fermi-surface topology in the CVC effect, we demonstrate the effect of anisotropy of electron pockets in the iron-pnictide case in Fig.3. Case A is the same as that in Fig.1. With just shrinking the Fermi-surface volume (case C), $\lambda(0)$ is enhanced due to increase of the nesting property, although the contribution of the CVC (the difference between solid and dashed lines in Fig.3) is the same degree. In contrast, in much anisotropic case B, the contribution of CVC is much suppressed, although w/o CVC is almost the same. Recently, in BaFe$_2$(As$_{1-x}$P$_x$)$_2$, it has been shown that the effective mass estimated by $\lambda^2(0)$, specific heat, and dHvA effect is quantitatively consistent within experimental error~\cite{rf:Walmsley2013}. This corresponds to the case B with small contribution of CVC. Indeed electron sheets of BaFe$_2$(As$_{1-x}$P$_x$)$_2$ is anisotropic rather than isotropic in the electronic band structure. Furthermore, the uniform mass enhancement over the Fermi surface discussed there is consistent with the isotropic suppression of $j^*_x({\bm k})$ in Fig.2 due to relatively large $\sqrt{\eta/A}$ as compared with the Fermi-surface volume.

Generally, the CVC term has more or less finite contribution on $\lambda(0)$, therefore the magnetic penetration depth can be sensitive to a presence of AF QCP, as compared with mass enhancement observed in specific heat and dHvA measurement. Observation of a peak- or cusp-like feature of $\lambda(0)$ provides a direct evidence of AF QCP in the superconducting state, while a finite jump indicates a first-order quantum phase transition. Note that the enhancement of $\lambda(0)$ is independent of the superconducting gap structure, nodal or nodeless. The zero-temperature penetration depth $\lambda(0)$ is always enhanced in a close proximity of the AF QCP even for a fully-gapped s-wave state, therefore $\lambda(0)$ can be a powerful tool to detect a QCP beneath the superconducting dome.
 
\begin{figure}[h]
\centering
\includegraphics[width=60mm]{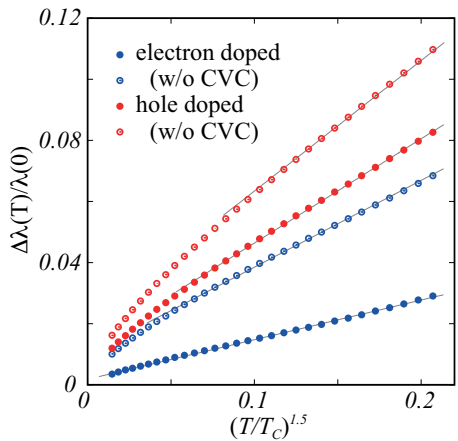}
\caption{(Color online) Magnetic penetration depth $\Delta\lambda(T)/\lambda(0)$ as a function of $(T/T_c)^{1.5}$. In the electron-doped case, anomalous power law $T^{1.5}$ can be observed over a wide temperature range. Solid lines denote $T^{1.5}$ dependence.}
\end{figure}

{\it Anomalous temperature dependence ---}
Next we discuss the $T$ dependence of $\lambda(T)$ in line-nodal superconductors near the AF QCP. In the ordinary nodal d-wave superconductors, $\Delta\lambda(T)=\lambda(T)-\lambda(0)$ should show $T$-linear behavior via the $T$ dependence of Yosida function $Y(\bm{k};T)$, which dominates low-energy nodal excitations. Near the AF QCP, the quasi-particle interactions $f_{\bm{k,k'}}$ have a sizable effect, and then a repetition of $Y(\bm{k};T)$ in Eq.(\ref{lambda}) can provide additional $T$ dependence; mainly the Maki-Thompson term $f_{\bm{k,k'}}Y(\bm{k};T)Y(\bm{k'};T) \sim T\times T=T^2$ can be enhanced. In Fig.4, we demonstrate $\Delta\lambda(T)/\lambda(0)$ as a function of $(T/T_c)^{1.5}$ with $\eta=0.03$ fixed in the cuprates case, where $\Delta({\bm k};T)=\Delta(T)(\cos k_x -\cos k_y)$ with the BCS-like $T$ dependence $\Delta(T)=\tanh \Bigl((\pi/2)\sqrt{T_c/T-1}\Bigr)$. 
We find anomalous $T^{1.5}$ dependence over a wide temperature range in the electron-doped case, while conventional $T$-linear behavior is found at low $T$ region in the hole-doped case and without the CVC. This means that in the CVC, a distance between nodal points and hot spots is the key parameter. In the hole-doped case, hot spots is far from the nodal points, and then low-energy excitation arises from an ordinary nodal excitations via $Y(\bm{k};T)$. On the other hand, in the electron-doped case, hot spots are located near the nodal points, and then nodal excitations are strongly affected by the CVC, mainly through the above-mentioned Maki-Thompson term. Thus anomalous $T^{1.5}$ behavior appears as a crossover from $T$ to $T^2$, and simultaneously the $T$-linear region shrinks (indeed invisible in the electron-doped case of Fig.4). As approaching the AF QCP, such anomalous power law becomes remarkable, since contribution of the CVC becomes crucially important. Note that in this mechanism $\Delta\lambda(T)$ remains strictly $T$-linear at $T \to 0$ limit, in sharp contrast to the interpolation formula, where $\Delta\lambda(T) \propto T^2$ rather at $T \to 0$ limit. Experimentally, there is no systematic study of $\lambda(T)$ in the electron-doped cuprates, but in the organic superconductors $\kappa-$(BEDT-TTF)$_2$X, an in-plane penetration depth indicates a crossover from $T^{1.5}$ to lower exponent with lowering $T$, which is well consistent with the present mechanism. Moreover, in the heavy-fermion superconductors, such as CeCoIn$_5$ and Ce$_2$PdIn$_8$, $T^n$ with $n=1.2\sim 1.5$ is observed over a wide $T$ range. These compounds possess complicated multi-Fermi surface with 3-dimensional corrugation. Then it is likely that some hot spots are located close to nodal points, which is just like the electron-doped case. In addition, inevitable impurities and/or lattice imperfection mask the $T$-linear behavior at $T \to 0$ limit. Thus, it can be generally expected that a power law $T^n$ with $n>1$ is observed in line-nodal superconductors that contain hot spots near the nodal points.

Finally, we comment on the iron-pnictide case of BaFe$_2$(As$_{1-x}$P$_x$)$_2$. In this system, the dominant spin fluctuation is an interband scattering between electron and hole sheets. The superconducting gap structure is still controversial; (a) horizontal nodes on the hole sheet~\cite{rf:Graser2010,rf:Suzuki2011,rf:Zhang2012} or (b) loop nodes on the electron sheet~\cite{rf:Mazin2010,rf:Yamashita2011,rf:Yoshida2013,rf:Saito2013}. In any case, two Yosida functions contained in the Maki-Thompson type contribution respectively originate from two different bands, electron sheet and hole sheet. If the gap structure on either band is nodeless fully-gapped, then its $Y(\bm{k};T)$ decays exponentially. In this case, the present mechanism does not work without the help of impurity scattering. However, it should be noted that what stabilizes such nodal structure in this system is not the above interband scattering, but the intraband scattering, between hole sheets in (a) or electron sheets in (b). If this scattering process is sufficiently strong, the quasi-particle interaction $f_{\bm{k,k'}}$ should contain this process, and then would lead to the anomalous $T^{1.5}$ behavior. Otherwise, we need another possible scenario, such as $T$ dependence of $f_{\bm{k,k'}}$ and $z_{\bm{k}}$ [\onlinecite{rf:Hashimoto2013}], some impurity effects~\cite{rf:Vorontsov2009}, and the effect of multi-gap structure. To explore this point, further study is needed. 

In conclusion, we have investigated the effect of the AF critical fluctuations on the magnetic penetration depth. Through the effect of mass enhancement and current vertex corrections, zero-temperature penetration depth is always enhanced at the AF QCP irrespective of the gap structure, in particular, dramatically in the iron-pnictide like case due to the Fermi-surface topology. Moreover, we find anomalous $T^{1.5}$ behavior over a wide temperature range even without impurities in line-nodal superconductors with hot spots near the nodal points. Thus we emphasize again that the magnetic penetration depth is a powerful tool to detect the AF QCP beneath the superconducting dome.

We thank Y. Matsuda and T. Shibauchi for stimulating our interest in this subject. We are also grateful to K. Ishida, Y. Yanase, S. Fujimoto, and K. Miyake for helpful discussion.
This work is supported by a Grant-in-Aid for Scientific Research on Priority Areas (No. 24540369, No. 23340095) from the Ministry of Education, Culture, Sports, Science and Technology, Japan.


\begin{thebibliography}{99}
\bibitem{rf:note01}
Note that $\lambda(0)$ in Eq.(\ref{lambda0}) is equivalent to the Drude weight, since the superfluid density becomes the electron density itself at $T=0$. This implies that the enhancement observed in $\lambda(0)$ does not strongly depend on gap structures.
\bibitem{rf:note02} 
Magnitude of the interaction $U$ is taken as the AF spin susceptibility does not diverge, referring to the FLEX results.
In the cuprate case, $U=1.0$eV is the on-site Coulomb repulsion, and $\alpha=1.5$. In the iron-pnictide case, $U=0.45$eV is the interband Coulomb interaction between hole and electron sheets, and $\alpha=1$ by neglecting the interband longitudinal fluctuation for simplicity.
\bibitem{rf:note04}
In the cuprates case, $\varepsilon_{\bm k} = -2t(\cos k_x+\cos k_y) + 4t'\cos k_x \cos k_y -2t''(\cos 2k_x+\cos 2k_y)$ with $t=0.5$eV and $t'/t=1/6$, and $t''/t=1/5$ [\onlinecite{rf:Yanase2003}]. The chemical potentials $\mu=-0.5t$ for the hole-doped case, and $\mu=0.5t$ for the electron-doped case.
For the iron-pnictide case, we apply simple two band model; $\varepsilon^{\rm h}_{\bm k} = -2t(\cos k_x+\cos k_y)-V$ for the hole band, and $\varepsilon^{\rm e}_{\bm k} = -2t'(\cos k_x+\cos k_y)-t''\cos(2k_x-2k_y)+V$ for the electron band. Here, $t=0.3$eV, $t'/t=2$, $t''/t=0$, $V/t=4.6$ and $\mu=-1.1t$, which were determined by considering the Fermi velocity and the Fermi-surface topology in the LDA band structure of BaFe$_2$(As$_{1-x}$P$_x$)$_2$ system~\cite{rf:Miyake2010}.
In this system, since the interband AF spin fluctuation is dominant, we neglect the intra-band scattering, and consider only the inter-band scattering for simplicity. In this case, we can use the single-band formalism as it is (do not need any band indices), since we can easily distinguish two bands from only positions of $\bm{k}$ in the first Brillouin zone.
\end{thebibliography}
\end{document}